\documentclass[aps,prl,twocolumn]{revtex4}

\usepackage{graphicx}
\usepackage{amssymb}

\begin{document} 

\title{Meta-material photonic funnels for sub-diffraction light compression and propagation}

\author{Alexander A. Govyadinov}
\affiliation{Physics Department, Oregon State University, 301 Weniger Hall, Corvallis, OR 97331, USA}
\author{Viktor A. Podolskiy}
\email{viktor.podolskiy@physics.oregonstate.edu}
\affiliation{Physics Department, Oregon State University, 301 Weniger Hall, Corvallis, OR 97331, USA}

\begin{abstract}
We present waveguides with photonic crystal cores, supporting energy propagation in subwavelength regions with a mode structure similar to that in telecom fibers. We design meta-materials for near-, mid-, and far-IR frequencies, and demonstrate efficient energy transfer to and from regions smaller than 1/25-th of the wavelength. Both positive- and negative-refractive index light transmissions are shown. Our approach, although demonstrated here in circular waveguides for some specific frequencies, is easily scalable from optical to IR to THz frequency ranges, and can be realized in a variety of waveguide geometries. Our design may be used for ultra high-density energy focusing, nm-resolution sensing, near-field microscopy, and high-speed all-optical computing.
\end{abstract}

\maketitle

While light emission by atoms, molecules, quantum wells, quantum dots and other quantum objects occurs from regions smaller than $10^{-8}\; m$-wide, light propagation takes place on much larger, wavelength scales, ranging from $10^{-2}\; m$ for GHz radar radiation to $10^{-6}\; m$ for optical light. Such a huge {\it scale difference}, effectively introduces a fundamental barrier between one's ability to generate light and to subsequently guide this light. Inefficient cross-scale coupling limits the resolution and sensitivity of near-field microscopes\cite{natureSNOM,sciNF}, prevents fabrication of ultra-compact all-optical processing circuits\cite{kivsharTrans,optTransOL}, integrated optoelectronic devices\cite{walba,lipsonNat} and other photonic systems. Here we present a class of compact waveguides capable of effective radiation transfer to, from, and in subwavelength regions. Our approach, although demonstrated here in circular waveguides for some specific frequencies, is easily scalable from optical to IR to THz frequency ranges, and can be realized in a variety of waveguide geometries.

\begin{figure}[thb]
\centerline{\includegraphics[width=8cm]{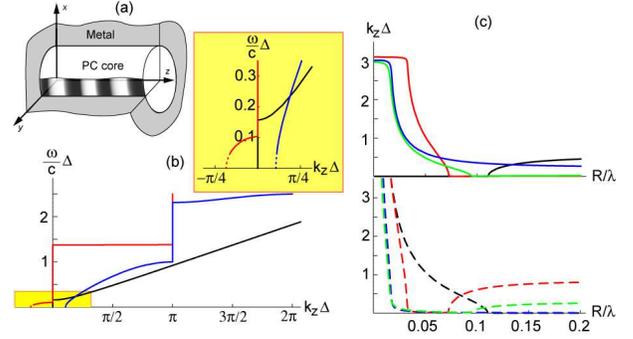}}
\caption{\label{figConfig} (a) Schematic geometry of a waveguide with PC core. (b) Unfolded dispersion diagram of the fundamental mode in three different photonic crystals (normalized to crystal period $\Delta$). Black line: TM mode in a waveguide with homogeneous core ($\epsilon_1=\epsilon_2;\;a_1=a_2$); note the frequency cut-off at $|k_z\Delta|\ll 1$ corresponding to diffraction limit; this behavior is similar to that of TE modes in PC structures described in the text. Blue line: TM mode with positive refractive index in PC with $\epsilon_1=12; \epsilon_2=-3.8;\; a_1=a_2=\Delta/2$. Red line: TM mode with negative refractive index in PC with $\epsilon_1=12; \epsilon_2=-73; a_1=a_2=\Delta/2$. Our design relies on $|k_z \Delta|\ll 1$ region (highlighted area and the inset in (b)), which is weakly affected by crystal disorder or other imperfections. The low-frequency regime (dotted region in the inset) is not realizable in nature since $\epsilon(\omega\rightarrow 0)$ has to be positive\cite{landauECM}.
(c) Confined mode propagation in PC waveguides. Absolute values of real (solid lines, upper graph) and imaginary (dashed lines, lower graph) parts of the wavevector are shown for TM01 mode in the Si-core (black), and Ag-Si (red), Si-SiC (blue), and InGaAs-AlInAs (green) PC structures (see text). The wavevector and radius are normalized to PC period and free-space wavelength respectively. Note that the negative-refraction modes propagate only in sufficiently thin waveguides\cite{podolskiyPRB}.
}
\end{figure}

A typical waveguide structure comprises at least two different regions: the transparent core that supports energy propagation, and the cladding that is used to prevent guided light from escaping the core. Although the design of both core and cladding may vary among different waveguides\cite{jonnapoulos,PBGRussel,omniguide,siegman,keilman02,stockmanPRL,atwaterNature,landauECM,BozhevolnyiPRL}, the wave propagation in the confined spaces has some universal, design-independent properties. Specifically, the electromagnetic radiation in any waveguide forms a series of system-specific waves: {\it waveguide modes}. The wavevector of each mode in the direction of mode propagation $k_z$ is related to the frequency through the following dispersion relation\cite{podolskiyPRB,PBGRussel,stockmanPRL}: 
\begin{equation}
\label{dispIso}
k_z^2 = \epsilon\nu\frac{\omega ^2}{c ^2},  
\end{equation}
where $\epsilon$ and $\nu$ are mode-specific propagation constants. In waveguides with non-magnetic isotropic homogeneous core $\epsilon$ is the dielectric permittivity of the core, and the parameter $\nu=1-\frac{\pi^2 m^2 c^2}{\epsilon R^2\omega^2}$ relates the frequency $\omega$, the speed of light in the vacuum $c$, the mode confinement radius $R$, and a generally non-integer mode number $m$ ($|m|\gtrsim 1$). The phase velocity of the mode is given by the effective index of refraction $n=\pm\sqrt{\epsilon\nu}$. 

Since the mode propagation is possible only when the effective refractive index is real, the product of $\epsilon\nu$ is required to be positive for such a propagation to take place. For the isotropic homogeneous system, the condition $n^2>0$ is equivalent to $\epsilon>0, \nu>0; n>0$. Thus, there exists a minimal critical radius of a waveguide supporting at least one confined propagating mode, $R_0\simeq \pi c/(\omega \sqrt{\epsilon})= \lambda/(2 \sqrt{\epsilon})$. The systems with radius $R<R_0$, formally described by $\epsilon>0,\nu<0$, reflect almost all incoming free-space-like radiation\cite{natureSNOM,footnoteEvWaves}, and are not suitable for energy transfer into subwavelength areas. 

The properties of the waveguide modes can be controlled by either changing the mode structure (modifying the parameter $m$), or by varying the waveguide core material (modifying the parameter $\epsilon$). Since there exists only a limited control over the dielectric permittivity of natural materials, the majority of modern subwavelength waveguides implement the former technique, and make use of the special type of modes at the metal-dielectric interfaces, known as surface waves, which formally correspond to $m^2<0$. Although these modes may in principle be used to transfer radiation to nano-scale\cite{stockmanPRL,atwaterNature,ZhangSci,BozhevolnyiPRL,keilman02,novotny,joannopolosPlasm,atwaterPRB}, their spatial structure is fundamentally different from that in free-space waves and telecom fiber modes. This structural difference requires non-trivial coupling mechanisms to convert the radiation from free-space to surface waves, typically associated with substantial coupling losses\cite{landauECM,novotny}. 

Here we present an alternative approach to compress and propagate the radiation below the free-space diffraction limit. Instead of changing the structure of the modes, we propose to change the waveguide itself. We use a {\it periodic array} of thin dielectric ($\epsilon>0$) and ``metallic'' ($\epsilon<0$) layers, widely known as a 1D photonic crystal (PC) medium, as a {\it meta-material waveguide core} (Fig.1a). 

In the case of PC layers perpendicular to the direction of mode propagation considered here, all modes of the system can be separated into two fundamentally different groups. The modes in the first group, known as TE waves, have their electric ($\mathbf{E}$) vector parallel to the layers, the modes in the second group (TM waves) have their magnetic ($\mathbf{H}$) vector parallel to the layers. Similar to the case of a homogeneous waveguide described above, the frequency and wavevector of a wave in a PC-core fiber can be related through the dispersion relation: 
\begin{eqnarray}
\nonumber
\cos [k_z (a_1+a_2)] = \cos (k_1 a_1) \cos (k_2 a_2)-
\\ \label{dispPhot}
\gamma \sin (k_1 a_1) \sin (k_2 a_2),
\end{eqnarray}
where $a_1$ and $a_2$ are thicknesses of the layers in the photonic crystal, $\epsilon_1>0$, and $\epsilon_2<0$ are their permittivities, $k_1^2=\epsilon_1\omega^2/c^2-\pi^2 m^2/R^2$, $k_2^2=\epsilon_2\omega^2/c^2-\pi^2 m^2/R^2$, and the parameter $\gamma$ is equal to $\gamma_{TE}=\frac{1}{2} (\frac{k_1}{k_2}+ \frac{k_2}{k_1})$ and $\gamma_{TM} = \frac{1}{2} (\frac{\epsilon_2}{\epsilon_1} \frac{k_1}{k_2}+ \frac{\epsilon_1}{\epsilon_2} \frac{ k_2}{k_1})$ for TE and TM modes respectively\cite{yariv}. The properties of several modes in typical PC systems are illustrated in Fig.~1. 

In the case when the period of the system is much smaller than the wavelength and the waveguide radius ($\vert k_1 a_1 \vert \ll 1$, $\vert k_2 a_2 \vert \ll 1$ and $\vert k_z (a_1 + a_2) \vert \ll 1$), Eq.~(\ref{dispPhot}) becomes identical to Eq.~(\ref{dispIso}) with polarization-specific propagation parameters $\epsilon$ and $\nu$, given by
\begin{eqnarray}
\epsilon=\epsilon^{TM}_{PC}=\epsilon^{TE}_{PC}=\frac{a_1\epsilon_1+a_2\epsilon_2}{a_1+a_2},
\nonumber
\\
\nu^{TM}_{PC}=1-\frac{ a_1 \epsilon_2 + a_2 \epsilon_1}
{ \epsilon_1 \epsilon_2 (a_1+a_2) }\frac{\pi^2 m^2 c^2}{R^2 \omega^2}, 
\label{eqEMT}
\\
\nu^{TE}_{PC}=1-\frac{1}{\epsilon_{PC}^{TE}}\frac{\pi^2 m^2 c^2}{ R^2 \omega^2}. 
\nonumber
\end{eqnarray}
In a way, the PC core plays the role of a homogeneous but anisotropic uniaxial meta-material with its optical axis parallel to the direction of mode propagation\cite{podolskiyJOSA}. The existence of propagating modes in these systems can be once again related to the effective index of refraction $n$. The propagation of TE modes is completely analogous to the wave propagation in isotropic systems described earlier. In contrast to this behavior, the PC structure can support TM waves in two fundamentally different regimes. The first regime, described by $\epsilon^{TM}_{PC}>0; \nu^{TM}_{PC}>0$ corresponds to positive index of refraction, while the second one, $\protect{\epsilon^{TM}_{PC}<0;} \protect{\nu^{TM}_{PC}<0}$ describes a negative refraction case\cite{Veselago,pendry,podolskiyPRB,ShelbySmithSci,soukoulisSci}, unachievable in conventional fiber- and plasmonic- waveguides\cite{stockmanPRL,sciNF,PBGRussel,joannopolosPlasm,BozhevolnyiPRL,atwaterNature,novotny,keilman02}. 

Both $n>0$ and $n<0$ structures may support wave propagation in highly-confined areas (Fig.~1). Indeed, the effective refractive index of a propagating TM mode in substantially thin ($R\ll\lambda$) strongly anisotropic system is inversely proportional to a mode confinement scale $R$. The decrease in $R$ is accompanied by a decrease of ``internal'' wavelength $\lambda/|n|$, virtually eliminating the diffraction limit in proposed structures. 

Furthermore, the PC waveguides with different refractive indices can be combined together, opening the door for the effective phase manipulation of light propagating in highly-confined areas. The possibility of such a versatile light management on nanoscale is one of the main points of this work. 

The self-adjustment of PC waveguide modes to the waveguide size, accompanied by compatibility between the mode structure in PC waveguides, telecom fibers, and free-space makes the PC systems ideal candidates for effective energy transfer between macroscopic wave-propagation systems and nano-scale objects. In these coupling structures, called {\it photonic funnels}, the size of the PC waveguides gradually varies along the direction of mode propagation, squeezing the light into $nm$-sized areas much like a conventional funnel squeezes water into a thin bottleneck, which is another main point of this work. 

The efficiency of energy compression in photonic funnels can be related to adiabaticity parameter $\delta=\left|\frac{d(1/k_z)}{dz}\right|$\cite{landauECM,stockmanPRL}, that defines the reflection by the funnel structure, and absorption in the system. Increase of the funnel's length typically reduces reflection loss, but increases absorption.

\begin{figure}[tbh]
\centerline{\includegraphics[width=4cm]{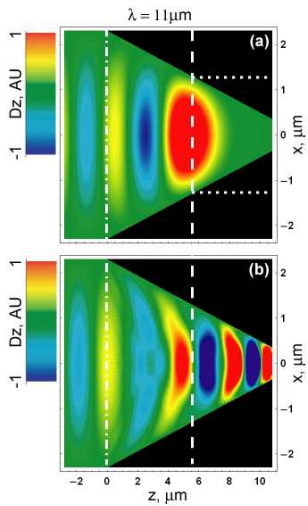}}
\caption{\label{figCompare} TM01 mode propagation ($\lambda=11\mu m$) from cylindrical Si waveguide ($z<0$) to a conical one with circular core; (a)~the wave in a system with Si core. Note the reflection from the point where radius reaches cut-off value $R_0\simeq 1.2 \mu m$ (dashed and dotted lines). Only $10^{-10}$ of energy is transmitted from $R=2.3\mu m$ to $R=0.35\mu m\sim\lambda/31$. This behavior is similar to that in tips of near-field microscopes\cite{natureSNOM,sciNF}. (b)~field concentration in Si-SiC PC funnel described in the text: $13\%$ of energy is transmitted to $R=0.35\mu m$, $16\%$ is reflected back to Si ($z<0$) waveguide. Note the dependence of the internal wavelength on the radius.
}
\end{figure}

\begin{figure}[tbh]
\centerline{\includegraphics[width=8cm]{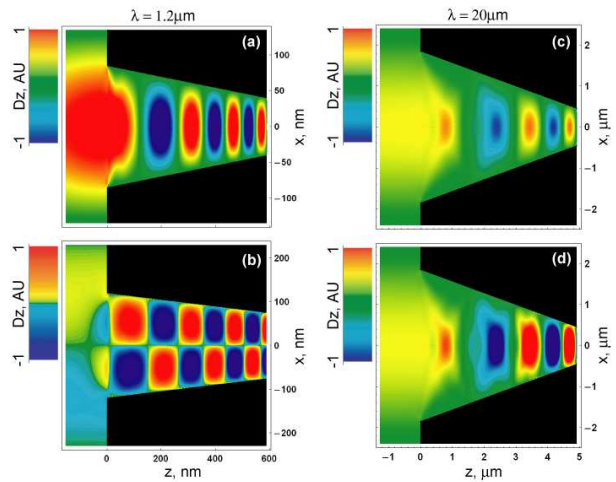}}
\caption{\label{figNIMs} Negative refractive index systems. (a-b) homogeneous Si waveguide ($z<0$) coupled to Ag-Si PC core cone as described in the text; $\lambda=1.2 \mu m$, (a) TM01 mode transfer from Si to a PC structure: $24\%$ of energy is transmitted from $R=135 nm$ to $R=40nm\sim\lambda/26$, $21\%$ is reflected back to $Si$ waveguide. (b) TM11 mode transfer from PC system ($z>0$) to Si waveguide: $11\%$ of energy is transferred from $R=75 nm \sim\lambda/16$ to $R=230 nm$, $13\%$ is reflected back. (c-d) TM01 energy transfer from AlInAs waveguide ($z<0$) to passive (c) and active (d) InGaAs-AlInAs PC-core; $\lambda=20\mu m$. The passive structure transmits $6\%$ of radiation from $R=2.4\mu m$ to $R=0.45\mu m \sim\lambda/44$. The material gain in active system ($\epsilon_{AlInAs}\simeq 10-i$) compensates for losses in the funnel structure and yields energy in $R=0.45\mu m$ to be $112\%$ of the incident one. Similar to Fig. 2, the internal wavelength is proportional to $R$}
\end{figure}

We now illustrate the light propagation in PC structures described here. Although our design does not impose any limitations on the waveguide geometry or waveguide boundary material, here we use conical waveguides with perfectly conducting metallic walls (see \cite{metalWalls}) and circular cores, having adiabaticity parameters $\delta\sim 0.1\ldots 0.3$, and defer the optimization of a photonic funnel geometry to future work. To demonstrate scalability of our design we calculate light propagation through the PC formed by (i)~ 100-nm-thick layers of SiC and Si with operating wavelength $\lambda=11\mu m$, (ii)~15-nm-thick layers of Ag and Si with operating wavelength $\lambda=1.2 \mu m$, and (iii)~$75\; nm$-thick InGaAs layers doped with electrons to $10^{19}\;{\rm cm}^{-3}$, and $150\; nm$ AlInAs barriers with operating wavelength $\lambda=20 \mu m$. (Fabrication of these structures is accessible through standard MOCVD, e-beam writing, or MBE techniques; see Refs.~\cite{Gmachl,ZhangSci,atwaterNature,lepeshkinPC} and references therein). 

To compute the light propagation in conical structures, we represent each structure by an array of cylindrical segments (in a way, this approach also accounts for the effects related to finite roughness of waveguide walls, unavoidable in any experiment). The typical radius step in our calculations is $10^{-3}\lambda$. We then represent the field in each cylindrical segment as a series of modes of a circular waveguide. We use Eqs.(\ref{dispPhot},\ref{eqEMT}) to calculate a mode propagation in each segment. In these calculations we use experimental values of permittivities for Ag, Si, SiC, and AlGaAs\cite{palik} and use Drude approach to describe InGaAs\cite{landauECM}. Finally, we use the boundary conditions to relate the modes in one segment to the modes in the neighboring segment, solving the problem of wave propagation through a photonic funnel. 

In Fig.2 we demonstrate the perspectives of photonic funnels by comparing the energy propagation through mid-IR PC waveguide with positive refraction described above to the propagation through the Si-core structure with identical geometry\cite{footnoteSiTheBest}. As expected, despite almost adiabatic radius compression, the energy in Si-core system reflects from the point corresponding to the cut-off radius of about $1.2\mu m$. In contrast to this behavior, PC system effectively compresses energy, and the propagation in the structure with radius as small as $0.35 \mu m\simeq\lambda/30$ is clearly seen. This PC provides a solution to the fundamental problem of coupling to the subwavelength domain, and allows transferring $13\%$ of energy, which is $10^{9}$ times better than its Si counterpart. 

The effective energy transfer across multiple scales in ``negative-refraction'' near- and far-IR PC systems is shown in Fig.3. Our calculations suggest that Ag-Si system may be used to transfer more than $20\%$ of energy to near-field zone. We expect that this number can be further optimized by reducing the wave reflection (currently $21\%$). 

The performance of the PC-based waveguides is limited by the PC microstructure, and by material absorption. The former introduces implicit inhomogeneity scale (PC period), where the ``effective medium'' approximation [Eq.~(\ref{eqEMT})] breaks down. The spatial dispersion, associated with field inhomogeneities on such a scale, leads to the mode cut-off and prohibits the mode propagation when the radius of a waveguide becomes smaller than PC period. The appearance of such a cut-off is shown in Fig.~1c. 

Material losses, on the other hand, lead to energy attenuation and limit the length of passive photonic funnels to $\sim 10\lambda$ which is acceptable for the majority of applications of these systems: near-field tips, ultra-compact detectors, wires in all-optical circuits, etc. This limitation is not applicable to waveguides with active cores. Indeed, material absorption can be substantially reduced, eliminated, or even reversed by implementing a gain medium into $\epsilon>0$ regions of PC\cite{pendryGain}. We illustrate this approach in Fig.~3d by introducing moderate gain into AlInAs part of the far-IR structure, which can be realized via quantum cascade technology\cite{Gmachl,belyanin}. 

Finally, we note that operating frequency of the photonic funnels described here can be changed from optical, to near-IR, to far-IR, to THz domain by varying the PC composition and periodicity. The PC-based waveguides may be used in ultra-compact all-optical and electro-optical devices, near-field microscopy, and other applications requiring effective sub-diffraction and cross-scale energy transfer, as well as in a variety of nonlinear optical applications\cite{boydPC} in positive- and negative-index materials since the energy compression and corresponding enhancement of local field will result in the strong enhancement of nonlinear field moments. 

The authors would like to thank E. Mishchenko for fruitful discussions


\begin{thebibliography}{}
\bibitem{footnoteEvWaves} We note however that an exponentially small part of radiation can penetrate through the sub-critical waveguide of finite length in a manner similar to the light transmission through thin metallic film.
\bibitem{metalWalls}At IR frequencies considered here, Ag, Au, and Al have dielectric permittivities at least an order of magnitude larger than the effective permittivities of our systems, and therefore are adequately described by ``perfectly conducting metal'' boundary conditions. The quantitatively weak effect of finite wall conductance on mode structure and propagation can be treated using standard perturbation techniques\cite{podolskiyJOSA}.
\bibitem{footnoteSiTheBest} The larger the dielectric constant of a material, the smaller the cut-off radius of a corresponding waveguide. Thus, since Si has one of the largest dielectric constants of all transparent natural materials, isotropic Si-core waveguides, shown in our simulation in a sense represent the ``minimum achievable'' cut-off radius with current technology
\end{thebibliography}
\end{document}